\documentclass[twocolumn,A4]{article}
\usepackage[dvips]{graphics}
\usepackage{color}
\definecolor{gold}{rgb}{0.85,0.66,0}
\definecolor{dblue}{rgb}{0,0,0.8}
\usepackage{setspace}
\begin{document}
\onecolumn
\begin{center}
{\bf{\Large {\textcolor{gold}{Tuning of electron transport through 
a quantum wire: An exact study}}}}\\
~\\
{\textcolor{dblue}{Santanu K. Maiti}}$^{\dag,\ddag}$,\footnote{{\bf
Corresponding Author}: Santanu K. Maiti \\ 
$~$\hspace {0.45cm} Electronic mail: santanu.maiti@saha.ac.in} \\
~\\
{\em $^{\dag}$Theoretical Condensed Matter Physics Division,
Saha Institute of Nuclear Physics, \\
1/AF, Bidhannagar, Kolkata-700 064, India \\
$^{\ddag}$Department of Physics, Narasinha Dutt College,
129 Belilious Road, Howrah-711 101, India} \\
~\\
{\bf Abstract}
\end{center}
We explore electron transport properties in a quantum wire attached to two 
metallic electrodes. A simple tight-binding model is used to describe the
system and the coupling of the wire to the electrodes (source and drain)
is treated through Newns-Anderson chemisorption theory. In our present 
model, the site energies of the wire are characterized by the relation
$\epsilon_i=W\cos\left(i \lambda^{\nu}\pi\right)$ where $W$, $\lambda$,
$\nu$ are three positive numbers. For $\nu=0$, the threshold bias voltage
of electron conduction across the bridge can be controlled very nicely 
by tuning the strength of the potential $W$. On the other hand, for 
$\nu \ne 0$, the wire becomes an aperiodic one and quite interestingly 
we see that, for some special values of $\nu$, the system exhibits a 
{\em metal-insulator} transition which provides a significant feature 
in this particular study. Our numerical results may be useful for 
fabrication of efficient switching devices.
\vskip 1cm
\begin{flushleft}
\end{flushleft}

\newpage
\twocolumn

\section{Introduction}

Quantum transport in low-dimensional systems like quantum wires,$^{1-2}$ 
quantum wells and quantum dots$^{3-4}$ provides several novel features 
due to reduced system dimensionality and lateral quantum confinement. 
The geometrical sensitivity of such systems makes them truly unique in 
offering the possibility of studying quantum transport in a very tunable 
environment. At present, manufacturing of the devices using single molecule
or cluster of molecules become more widespread and they provide a 
signature in the design of future nano-electronic circuits where the 
electron transport is predominantly coherent.$^{5-6}$ Based on the 
pioneering work of Aviram and Ratner$^7$ in which a molecular electronic 
device has been predicted for the first time, the development of a 
theoretical description of molecular electronic devices has been 
pursued. Later, several experiments$^{8-10}$ have been carried out 
in different molecular bridges to understand the basic mechanisms 
underlying such transport. Though electron transport properties through 
several bridge systems have been investigated elaborately both 
theoretically as well as experimentally, yet the complete knowledge 
of conduction mechanism in this scale is not very well established 
even today. Several significant factors are there which control the 
electron transport across a bridge system, and all these effects have 
to be taken into account properly to characterize such transport. For
our illustration, here we describe very briefly some of these factors.
(i) Molecular coupling to the side attached electrodes and the
electron-electron correlation.$^{11}$ Understanding of the molecular
coupling to the electrodes under non-equilibrium condition is a major
challenge in this particular study. (ii) Geometry of the molecule 
itself. To emphasize it, Ernzerhof {\em et al.}$^{12}$ have predicted
several model calculations and provided some new interesting results.
(iii) Quantum interference effect$^{13-18}$ of electronic waves 
passing through the bridge system. (iv) Dynamical fluctuation in 
small-scale devices is another important factor which plays an 
active role and can be manifested through measurement of {\em shot 
noise}, a direct consequence of the quantization of charge. It can be 
used to obtain information on a system which is not available directly 
through conductance measurements, and is generally more sensitive 
to the effects of electron-electron correlations than the average 
conductance.$^{19-20}$ Beside these factors, several other parameters 
of the Hamiltonian that describe a system also provide significant 
effects in the determination of current through a bridge system.

In the present paper, we will investigate the electron transport 
properties of a quantum wire attached to two semi-infinite 
one-dimensional ($1$D) metallic electrodes. It is well established from 
the Bloch's theorem that, a one-dimensional quantum wire subject to 
the periodic on-site potentials, exhibits all extended eigenstates. 
While, for the random distribution of the site potentials, localized 
states are the only allowed solutions as predicted by the Anderson 
model.$^{21}$ Thus, for these two different class of materials we 
get either all extended states or localized states.$^{21-22}$ But 
there is a special type of quasi-random potentials ($\epsilon_i=W\cos\left(
i \lambda^{\nu} \pi \right)$) lying in between the above mentioned 
potential distributions which have attracted a lot of interest$^{23-34}$ 
in last few decays. Our numerical results show that for a special choice 
of the parameter $\nu$ ($\nu=0$), threshold bias voltage of electron 
conduction across the wire can be controlled in a tunable way by 
changing the strength of the potential $W$. While for the non-zero 
value of $\nu$, the wire becomes an aperiodic one and several
interesting results are obtained. Quite significantly we see that, 
for some particular values of the parameter $\nu$, the system shows
{\em metal-insulator} transition which clearly manifests the existence
of the {\em mobility edge} in conductance spectrum. In our present 
discussion, we use a simple tight-binding model to incorporate 
electron transport through a quantum wire, and adopt the Newns-Anderson 
chemisorption model$^{35-37}$ for the description of electrodes and for 
the interaction of electrodes to the wire. Our numerical studies might 
throw new light, both in the context of basic physics and possible 
technological applications.

The paper is organized as follows. In Section $2$, we describe the model 
and methodology for the calculation of transmission probability ($T$) 
and current ($I$) through a quantum wire attached to two metallic 
electrodes using single particle Green's function formalism. Section 
$3$ discusses the significant results, and finally, we summarize our 
study in Section $4$.

\section{Model and the theoretical description}

\begin{figure}[ht]
{\centering \resizebox*{7cm}{0.8cm}{\includegraphics{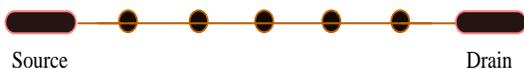}}\par}
\caption{Schematic view of a quantum wire attached to two semi-infinite 
$1$D electrodes. The filled black circles correspond to the atomic 
sites in the wire (for color illustration, see the web version).}
\label{wire}
\end{figure}

This section describes the model and technique for the calculation of 
transmission probability ($T$), conductance ($g$) and current ($I$) 
through a quantum wire attached to two $1$D metallic electrodes using 
the Green's function method. The schematic view of such a bridge 
system is illustrated in Fig.~\ref{wire}. In actual experimental set up, 
these two electrodes made from gold are used and the wire attached to 
them via thiol groups in the chemisorption technique and in making such 
contact, hydrogen (H) atoms of the thiol groups remove and the sulfur 
(S) atoms reside. 

For low bias voltage and temperature, conductance $g$ of the wire is 
determined from the Landauer conductance formula,$^{38-39}$
\begin{equation}
g=\frac{2e^2}{h} T
\label{equ1}
\end{equation}
where, the transmission probability $T$ can be written in the 
form,$^{38-39}$
\begin{equation}
T={\mbox{Tr}} \left[\Gamma_S G_w^r \Gamma_D G_w^a\right]
\label{equ2}
\end{equation}
where, $G_w^r$  and $G_w^a$ correspond to the retarded and advanced 
Green's functions of the wire, and $\Gamma_S$ and $\Gamma_D$ describe 
the coupling 
of the wire to the source and drain, respectively. The Green's function 
of the wire is expressed as,
\begin{equation}
G_{w}=\left(E-H_{w}-\Sigma_S-\Sigma_D\right)^{-1}
\label{equ3}
\end{equation}
where, $E$ is the energy of the injecting electron and $H_{w}$ 
corresponds to the Hamiltonian of the wire. Within the non-interacting 
picture this Hamiltonian can be written in the tight-binding model as,
\begin{equation}
H_{w}=\sum_i \epsilon_i c_i^{\dagger} c_i + \sum_{<ij>} t 
\left(c_i^{\dagger} c_j + c_j^{\dagger} c_i\right)
\label{equ4}
\end{equation}
In this expression, $c_i^{\dagger}$ ($c_i$) represents the creation 
(annihilation) operator of an electron at site $i$, $\epsilon_i$'s are the 
on-site energies and $t$ corresponds to the nearest-neighbor hopping strength. 
In our present study, the on-site potentials of the wire are described by 
the expression,
\begin{equation}
\epsilon_i = W \cos\left(i \lambda^{\nu} \pi \right)
\end{equation}
where, $W$ be the strength of the potential and the parameters $\lambda$ 
and $\nu$ are the two other positive numbers define the tight-binding 
problem. Depending on the value of the parameter $\nu$, the wire becomes
non-aperiodic ($\nu=0$) or aperiodic ($\nu \ne 0$), and both for these 
two different choices of $\nu$ we get several interesting results. In 
order to describe the side attached electrodes, viz, source and drain, we 
use the standard tight-binding Hamiltonian as prescribed in Eq.~(\ref{equ4})
and parametrized it by constant on-site potential $\epsilon_0$ and 
nearest-neighbor hopping integral $v$. The wire is coupled to these 
electrodes through the parameters $\tau_S$ and $\tau_D$,
where they (coupling parameters) correspond to the coupling strengths 
to the source and drain respectively. In Eq.~(\ref{equ3}), the 
parameters $\Sigma_S$ and $\Sigma_D$ correspond to the self-energies due 
to coupling of the wire to the source and drain, respectively. All the 
information of this coupling are included into these two self-energies 
and are described by the Newns-Anderson chemisorption model.$^{35-37}$ 
This Newns-Anderson model permits us to describe the conductance in terms 
of the effective wire properties multiplied by the effective state 
densities involving the coupling, and allows us to study directly the 
conductance as a function of the properties of the electronic structure 
of the wire between the electrodes.

The current passing through the wire can be regarded as a single electron
scattering process between the two reservoirs of charge carriers. The
current-voltage relationship can be obtained from the expression,$^{38}$
\begin{equation}
I(V)=\frac{2e}{h}\int \limits_{-\infty}^{\infty} 
\left(f_S-f_D\right) T(E) ~dE
\label{equ5}
\end{equation}
where $f_{S(D)}=f\left(E-\mu_{S(D)}\right)$ gives the Fermi distribution
function with the electrochemical potential $\mu_{S(D)}=E_F\pm eV/2$.
Usually, the electric field inside the wire, especially for small wires, 
seems to have a minimal effect on the $g$-$E$ characteristics. Thus 
it introduces very little error if we assume that, the entire voltage is
dropped across the wire-electrode interfaces. The $g$-$E$ characteristics
are not significantly altered. On the other hand, for larger system sizes
and higher bias voltage, the electric field inside the wire may play a
more significant role depending on the size and structure of the
wire,$^{40}$ though the effect is quite small.

All the results presented in this communication are worked out for zero
temperature. However, they should remain valid even in a certain range
of finite temperature ($\sim 300$ K). This is due to the fact that the 
broadening of energy levels of the wire due to the electrode-wire
coupling is, in general, much larger than that of the thermal 
broadening.$^{38}$ For simplicity, we take the unit $c=e=h=1$ in our 
present calculations.

\section{Numerical results and discussion}

This section demonstrates the transport properties of a quantum wire. 
Here we concentrate our results on clarifying the dependence of 
conductance and current on (i) the parameters $W$, $\lambda$ and 
$\nu$ and (ii) the coupling strength of the wire
\begin{figure}[ht]
{\centering \resizebox*{7.8cm}{10cm}{\includegraphics{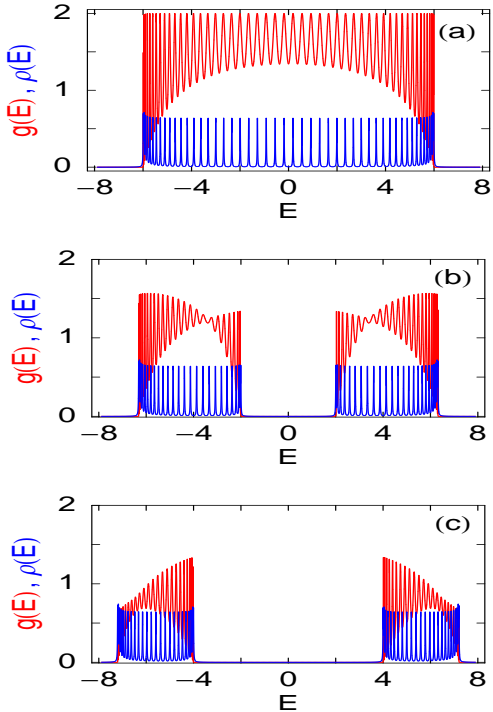}}\par}
\caption{$\nu=0$. $g$-$E$ (red color) and $\rho$-$E$ (blue color) curves
in the strong-coupling limit for some quantum wires with total number of 
sites $N=50$, where (a), (b) and (c) correspond to $W=0$, $2$
and $4$, respectively (for color illustration, see the web version).}
\label{cond1}
\end{figure}
to the two side attached electrodes. The parameter $\lambda$ is an
irrational number which is set to $(1+\sqrt{5})/2$, for our illustration.
Out of these three parameters ($W$, $\lambda$, $\nu$), $\nu$ is the
most significant one. For the three distinct regimes of $\nu$, we get
several important results which we will discuss here one by one. In the
first regime, $\nu=0$. Therefore, the on-site energy ($\epsilon_i$)
alternates between the two values $W$ and $-W$. Thus the wire becomes 
a non-aperiodic one. 
The other two regimes are $0<\nu< 1$ and $\nu > 1$, where we choose 
$\nu=0.5$ and $\nu=1.4$, respectively. In the same footing, the 
wire-to-electrode coupling strength has a strong dependence in the 
transport phenomena. To reveal this fact, here we focus our results 
for the two distinct regimes, the so-called weak- and strong-coupling 
regimes, respectively. These two regions are described by the conditions 
$\tau_{\{S,D\}} << t$ and $\tau_{\{S,D\}} \sim t$, respectively. The 
values of these parameters for the two distinct regimes are taken as: 
$\tau_S=\tau_D=0.5$, $t=3$ (weak-coupling) and $\tau_S=\tau_D=2.5$, 
$t=3$ (strong-coupling). Throughout the calculations, the on-site 
\begin{figure}[ht]
{\centering \resizebox*{7.8cm}{10cm}{\includegraphics{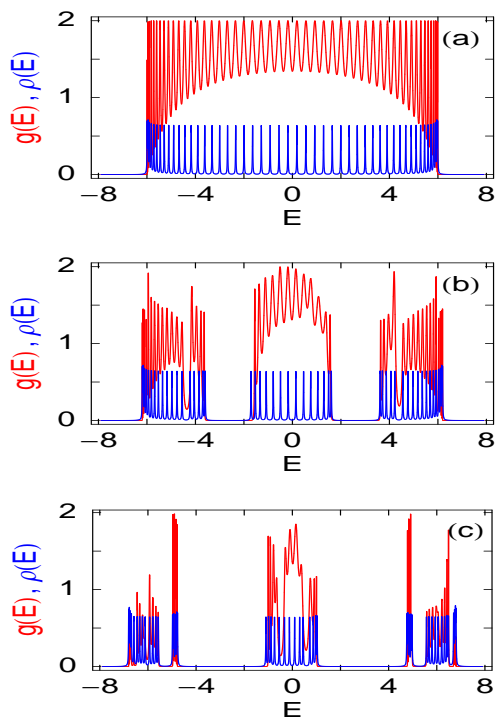}}\par}
\caption{$\nu=0.5$. $g$-$E$ (red color) and $\rho$-$E$ (blue color) curves
in the strong-coupling limit for some quantum wires with total number of 
sites $N=50$, where (a), (b) and (c) correspond to $W=0$, $2$
and $4$, respectively (for color illustration, see the web version).}
\label{cond2}
\end{figure}
potential and hopping integral in the electrodes are set as 
$\epsilon_0=0$ and $v=4$, respectively. The equilibrium Fermi energy
$E_F$ is fixed at $0$.

To illustrate our results, let us first concentrate to the regime 
where $\nu=0$. As representative examples, the characteristic properties 
of the conductance $g$ (red color) and density of states (DOS) $\rho$ 
(blue color) for some typical quantum wires are presented in 
Fig.~\ref{cond1}, where (a), (b) and (c) correspond to $W=0$, $2$ and 
$4$, respectively. For this particular value of $\nu$, the wire becomes 
a non-aperiodic one and the on-site energies become $W$ or $-W$ depending 
on whether the site index $i$ is even or odd. Figure~\ref{cond1}(a) 
represents the result for a completely perfect wire where all the site 
energies are set to zero (as $W=0$ for this particular case). The 
conductance shows sharp resonant peaks throughout the bandwidth of 
the wire ($-2t$ to $+2t$), associated with the energy eigenvalues of 
it. This is clearly observed by superposing the picture of the density 
of states on the conductance profile. At these resonances, the 
conductance gets the value $2$, and accordingly, 
\begin{figure}[ht]
{\centering \resizebox*{7.8cm}{10cm}{\includegraphics{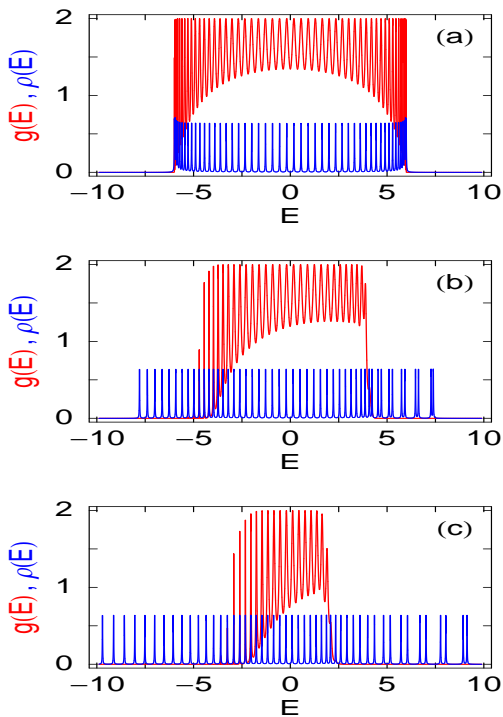}}\par}
\caption{$\nu=1.4$. $g$-$E$ (red color) and $\rho$-$E$ (blue color) curves
in the strong-coupling limit for some quantum wires with total number of 
sites $N=50$, where (a), (b) and (c) correspond to $W=0$, $2$
and $4$, respectively (for color illustration, see the web version).}
\label{cond3}
\end{figure}
the transmission probability $T$ becomes unity, since from the Landauer 
conductance formula the relation $g=2T$ holds (see Eq.~(\ref{equ1}) with 
$e=h=1$ in our present description). Thus the conductance spectrum
manifests itself the electronic structure of the system. In 
Figs.~\ref{cond1}(b) and (c), the results are shown for the finite
values of $W$, and quite interestingly we see that, a gap appears in the
conductance spectra across the energy $E=0$. This is due to the existence
of the two energy bands separated by twice the width of disorder. 
Therefore, the gap between the two bands increases with the increase 
of the strength $W$, which is clearly observed from these figures. 
Thus we can tune the electron conduction across the wire by controlling 
the parameter $W$. This provides an interesting phenomenon, and it can 
be much more clearly understood from our study of the current-voltage 
($I$-$V$) characteristics.

Now we consider the results for the aperiodic quantum wires where $\nu$
becomes finite. As illustrative examples, in Fig.~\ref{cond2}, we display
the conductance and density of states for some typical aperiodic
\begin{figure}[ht]
{\centering \resizebox*{7.8cm}{10cm}{\includegraphics{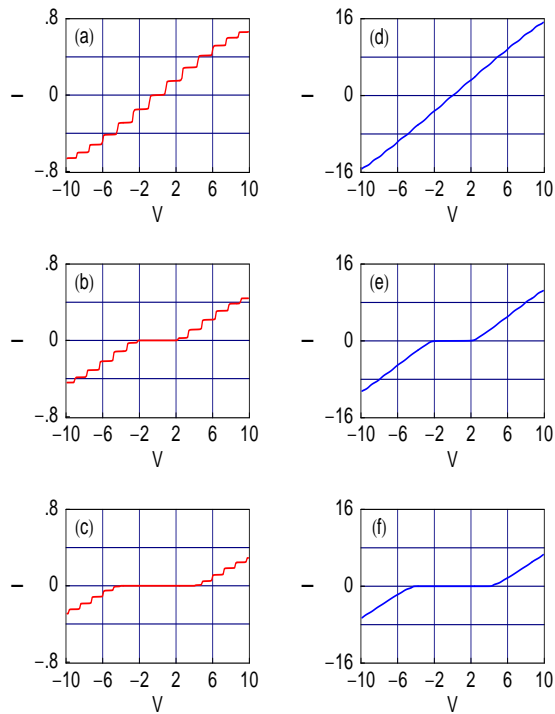}}\par}
\caption{$\nu=0$. $I$-$V$ curves in the weak- (red color) and 
strong-coupling (blue color) limits for some quantum wires with total 
number of sites $N=20$. The $1$st, $2$nd and $3$rd rows correspond to 
$W=0$, $1$ and $2$, respectively (for color illustration, see the web 
version).}
\label{curr1}
\end{figure}
quantum wires, where (a), (b) and (c) correspond to $W=0$, $2$ and
$4$, respectively. For these wires, we set $\nu=0.5$. When $W=0$, the 
result is exactly similar to that as given in Fig.~\ref{cond1}(a). 
We re-plot this result once again to compare the other results for 
the finite values of $W$ quite clearly. Our numerical results show
that, for this particular regime of $\nu$ ($\nu<1$), the conductance
spectrum breaks into three distinct regions associated with the
DOS spectrum.  
Quite significantly it is observed that, the energy separation between 
these bands increases with the strength of $W$. Thus we get three 
separate energy bands for these kind of aperiodic quantum wires. Such a 
behavior can be used to construct multiple switching devices, which is 
really a very interesting observation in the transport community. 

The most significant result is obtained when we choose $\nu=1.4$ i.e., 
in the regime $\nu>1$. As representative examples, we plot the $g$-$E$
and $\rho$-$E$ characteristics in Fig.~\ref{cond3} for some typical
aperiodic quantum wires subject to this particular value of $\nu$,
where (a), (b) and (c) correspond to $W=0$, $2$ and $4$, respectively.
Figure~\ref{cond3}(a) is identical with those as in Fig.~\ref{cond1}(a) 
and Fig.~\ref{cond2}(a). For non-zero values of $W$, Figs.~\ref{cond3}(b)
\begin{figure}[ht]
{\centering \resizebox*{7.8cm}{9.5cm}{\includegraphics{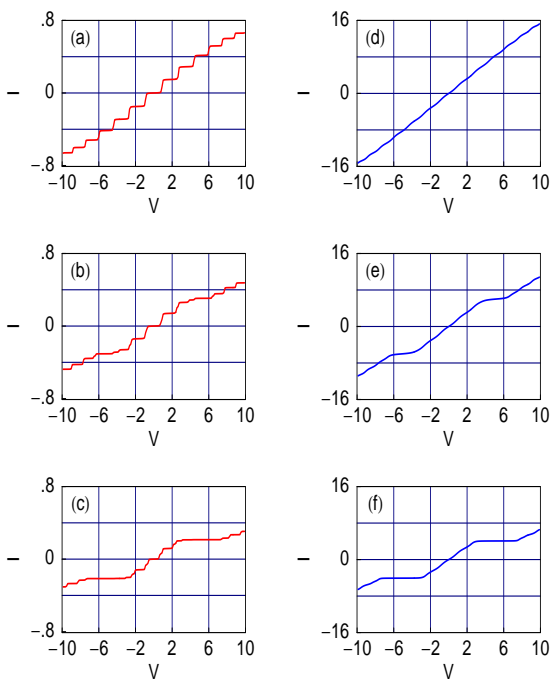}}\par}
\caption{$\nu=0.5$. $I$-$V$ curves in the weak- (red color) and 
strong-coupling (blue color) limits for some quantum wires with total 
number of sites $N=20$. The $1$st, $2$nd and $3$rd rows correspond to 
$W=0$, $1$ and $2$, respectively (for color illustration, see the web 
version).}
\label{curr2}
\end{figure}
and (c) clearly show that there are eigenstates existing in the energy
regimes for which the conductance is completely zero. This illustrates
the transition from the conducting (high $g$) to the non-conducting
or insulating phase, the so-called {\em metal-insulator} transition
in the system. Thus from the conductance and the DOS spectra, a
crossover from a completely opaque to a fully transmitting zone is easily
understood. In view of such a crossover, one can then set the Fermi energy
$E_F$ at a suitable energy zone in the spectrum and control the transmission
characteristics. This enhances the prospect of such quantum wires as 
novel switching devices.

Up to now, all the results studied here are described only for the 
strong-coupling limit. Quite similar feature will also be observed in the 
conductance spectra for the weak-coupling limit in these three distinct 
regimes of $\nu$. The only difference appears in the conductance spectra 
in these two coupling limits is that, in the weak-coupling case the
widths of the resonant peaks become much narrow than that of the
strong-coupling case. This broadening solely depends on the imaginary 
parts of the two self-energies $\Sigma_S$ and $\Sigma_D$.$^{38}$
Now for these quantum wires, since we get many resonant peaks, it is 
quite difficult to observe the difference between the widths of the 
resonant peaks in the conductance spectra for the two different coupling 
cases, and therefore, we do not show the results for the weak-coupling 
case. The effects of the electrode-wire coupling can be much more clearly 
understood from the current-voltage characteristics which we are going
to discuss in our forthcoming parts of this paper.

Current across the bridge is computed from the integration procedure
of the transmission function $T$ as prescribed in Eq.~(\ref{equ5}). The
transmission function varies exactly similar to that of the conductance
spectra as illustrated above, differ only in magnitude by the factor $2$
since the relation $g=2T$ holds from the Landauer conductance formula 
(Eq.~(\ref{equ1})). Let us first discuss the results for the case $\nu=0$.
As typical examples, in Fig.~\ref{curr1}, we plot the current-voltage 
($I$-$V$) characteristics for some non-aperiodic quantum wires ($N=20$),
where the $1$st, $2$nd and $3$rd rows correspond to $W=0$, $1$ and $2$,
respectively. The red lines represent the results for the weak-coupling
limit, while the blue lines denote the results in the limit of 
strong-coupling. Both for these two limiting cases several significant
results are observed. In the weak wire-electrode coupling, the current
exhibits staircase-like structure with fine steps as a function of the
applied bias voltage. This is due to the existence of the sharp resonant
peaks in the conductance spectra in this limit of coupling, since 
current is computed by the integration method of the transmission 
function $T$. With the increase of the applied bias voltage, 
electrochemical potentials on the electrodes are shifted gradually, and 
finally cross one of the quantized energy levels of the wire. Therefore, 
a current channel is opened up and the current-voltage characteristic 
curve provides a jump. On the other hand, for the strong wire-electrode
coupling, current varies almost continuously with the applied bias 
voltage and achieves much large amplitude than the weak-coupling case. 
This is because the resonant peaks get broadened due to the broadening 
of the energy levels in the strong-coupling limit which provide much 
larger current amplitude as we integrate the transmission function $T$
to get the current. Thus by tuning the wire-to-electrode coupling, one 
can achieve very high current from the very low one. Now the effect of 
\begin{figure}[ht]
{\centering \resizebox*{7.8cm}{9.5cm}{\includegraphics{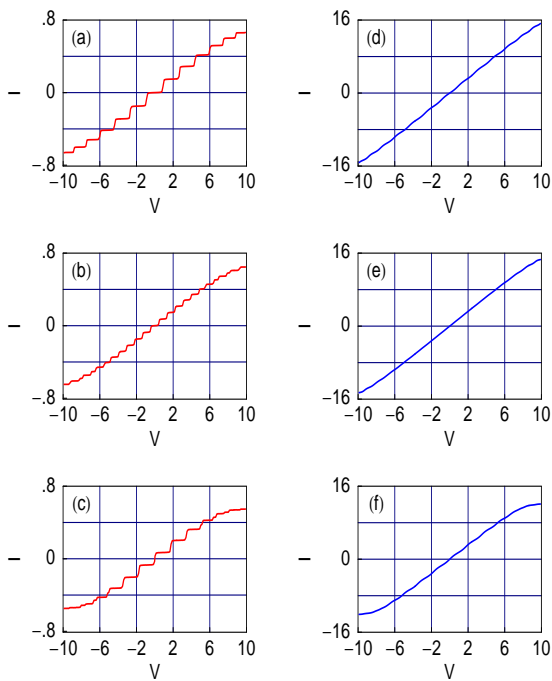}}\par}
\caption{$\nu=1.4$. $I$-$V$ curves in the weak- (red color) and 
strong-coupling (blue color) limits for some quantum wires with total 
number of sites $N=20$. The $1$st, $2$nd and $3$rd rows correspond to 
$W=0$, $1$ and $2$, respectively (for color illustration, see the web 
version).}
\label{curr3}
\end{figure}
$W$ is quite interesting in this particular study. For the non-zero 
value of $W$, current across the wire is obtained beyond some finite 
bias voltage (see Figs.~\ref{curr1}(b) and (e)), which is associated 
with the conductance and DOS spectra (Fig.~\ref{cond1}). This threshold 
bias voltage increases with the increase of the strength $W$ (see 
Figs.~\ref{curr1}(c) and (f)). Thus one can shift the threshold 
bias voltage gradually by changing the parameter $W$ accordingly.

Next we make an attention to the current-voltage characteristics for the
aperiodic quantum wires where we set $\nu=0.5$. In Fig.~\ref{curr2}, we
represent the current-voltage characteristics for some aperiodic quantum 
wires, where the $1$st, $2$nd and $3$rd rows correspond to $W=0$, $1$ and
$2$, respectively. The red and blue curves represent the identical meaning 
as in Fig.~\ref{curr1}. Both for the weak- and strong-coupling cases, we 
get quite similar features as in the $\nu=0$ case i.e, step-like in the
limit of weak-coupling and continuous-like in the other limiting case.
But the key behavior is that, here one gets the current across the wire
for much low bias voltage in the weak coupling case, while in the 
strong wire-electrode coupling the current is available as long as the 
bias voltage is applied. In the presence of $W$, an interesting behavior
is observed. Our results show that, for a wide range of the applied bias
voltage, the current remains constant and after that it again increases 
with the voltage $V$. This constant current region can also be controlled 
by tuning the appropriate parameters.
 
Finally, we consider the particular case where we fix $\nu=1.4$. The 
results are shown in Fig.~\ref{curr3}, where the $1$st, $2$nd and $3$rd
rows correspond to the similar strengths of $W$ as in the previous cases.
The red and blue curves also represent the identical meaning as above.
The coupling effects are same as described earlier. For this particular
value of $\nu$, the current is observed for any non-zero value of $W$
in both these two limiting cases as long as the bias voltage is applied
(see the $2$nd and $3$rd rows of Fig.~\ref{curr3}). Like in 
Fig.~\ref{curr2}, here we will also get the saturation region of the 
current that appears at quite higher values of the bias voltage $V$, 
which is not shown here in the figure. This saturation region continues 
with the bias voltage and there is no such possibility of getting the 
current further for the higher voltages as in the case of $\nu=0.5$.

\section{Concluding remarks}

In conclusion, the electron transport properties in a quantum wire 
attached to two semi-infinite $1$D metallic electrodes have been 
investigated by using the single particle Green's function formalism. 
We have used
a simple tight-binding model to describe the system where the coupling
of the wire to the electrodes has been treated through the Newns-Anderson 
chemisorption theory. In this article, we have described our results for
the three typical values of $\nu$ those are associated with the three 
different regimes of this parameter. For $\nu=0$, the wire becomes
a non-aperiodic one and in such a case, the threshold bias voltage of
electron conduction can be tuned significantly by controlling the strength
of the potential $W$. On the other hand, for the non-zero value of $\nu$, 
the wire becomes an aperiodic one and several other key features are 
obtained. In the regime $0<\nu<1$, conductance spectrum breaks into 
three separate bands associated with the energy spectrum. By tuning the
parameter $W$, we can change the gap between the bands in the conductance 
spectrum, which is responsible for the constant current across the wide
range of applied bias voltage. The most remarkable result is observed
in the regime $\nu>1$. In this regime ($\nu=1.4$), our numerical results 
have shown a cross over from the conducting (high $g$) to non-conducting
or insulating phase. This illustrates a {\em metal-insulator} transition 
in the system. In view of such a crossover, one can tune the Fermi energy
to a suitable energy zone in the spectrum, and thus, can be able to control
the transmission properties. This aspect may be utilized in designing a 
tailor made switching device. In this article we have also described the
effects of the wire-electrode coupling which have been clearly visible
from our current-voltage characteristics. The current shows sharp
staircase-like structure in the weak-coupling limit, while it varies 
quite continuously and achieves very high value in the limit of 
strong-coupling. Thus, we can emphasize that, care should be taken on the 
coupling effects in fabrication of electronic circuits by using nano-scale 
systems.

Throughout our discussions we have used several approximations by 
neglecting the effects of the electron-electron interaction, all the 
inelastic scattering processes, the Schottky effect, the static Stark 
effect, etc. Beside these, here we have also made another one 
approximation by calculating all the results at absolute zero 
temperature, and we have already stated about it at the end of Section 
$2$. Instead of calculating all the results at zero temperature, we 
can also determine these results at some finite temperatures. But, our 
presented results will not change significantly even for some finite 
non-zero temperatures. More studies are expected to take into account 
all these approximations for our further investigations. 

\vskip 0.3in
\noindent
{\bf\Large Acknowledgments}
\vskip 0.2in
\noindent
I acknowledge with deep sense of gratitude the illuminating comments and
suggestions I have received from Prof. Arunava Chakrabarti and Prof. 
Shreekantha Sil during the calculations.

\end{document}